\documentclass[12pt]{amsart}
\usepackage{cite}
\usepackage{epsfig}
\usepackage{courier}
\usepackage{amsmath,amssymb,amsthm,amscd,mathtools, graphicx, mathtext,color,wrapfig}
\usepackage[numbers,sort&compress]{natbib}

\makeatletter

\@addtoreset{equation}{section}
\makeatother

\newcommand{\beq}{\begin{equation}}
\newcommand{\eeq}{\end{equation}}

\newcommand{\cal}{\mathcal}

\theoremstyle{definition}

\begin{document}
\baselineskip=18pt  
\baselineskip 0.713cm

\begin{titlepage}

\setcounter{page}{0}

\renewcommand{\thefootnote}{\fnsymbol{footnote}}


\vskip 1.0cm

\begin{center}
{\LARGE \bf
Gauge/Vortex Duality and AGT
\\
\vskip  0.5cm
}

\vskip 0.5cm

{\large
Mina Aganagic$^{1,2}$ and Shamil Shakirov$^{1,2,3}$}
\\
\medskip

\vskip 0.5cm

{\it
$^1$Center for Theoretical Physics, University of California, Berkeley,  USA\\
$^2$Department of Mathematics, University of California, Berkeley,  USA\\
}

\end{center}

\vskip 0.5cm

\centerline{{\bf Abstract}}
\medskip
AGT correspondence relates a class of 4d gauge theories in four dimensions to conformal blocks of Liouville CFT. There is a simple proof of the correspondence when the conformal blocks admit a free field representation. In those cases, vortex defects of the gauge theory play a crucial role, extending the correspondence to a triality. This makes use of a duality between 4d gauge theories in a certain background, and the theories on their vortices. The gauge/vortex duality is a physical realization of large $N$ duality of topological string which was conjectured in \cite{DVt} to provide an explanation for AGT correspondence. This paper is a review of \cite{AHKS}, written for the special volume edited by J. Teschner.
\noindent\end{titlepage}
\setcounter{page}{1} 


\section{Introduction}

Large $N$ duality plays the central role in understanding dynamics of physical string theory. This duality is inherited by the simpler, topological string with target space a Calabi-Yau three-fold \cite{GV, DV, IH}. The topological large $N$ duality, like the large $N$ duality of the physical string theory, relates the gauge theory on D-branes to closed topological string on a different background. In the topological string case, the duality is in principle tractable, since topological string is tractable.

In some cases, study of topological string theory is related to studying supersymmetric gauge theory in 4d with ${\cal N}=2$ supersymmetry, see e.g. \cite{N2, Neitzke:2004ni} and [V:13]. It is natural to ask what the large $N$ duality of topological string theory means in gauge theory terms. We will see that the large $N$ duality of topological string becomes a gauge/vortex duality \cite{DH1, DH2, Simonstalk} which relates a 4d gauge theory in a variant of 2d $\Omega$ background with flux, and the theory living on its vortices.\footnote{For early studies leading to \cite{DH1, DH2, Simonstalk}, see \cite{Dorey, DHT, HananyTong, HananyTong2, Shifman:2004dr}. } The vortices in the gauge theory play the role of D-branes of the topological string. In fact, the gauge theory duality implies the topological string duality, but not the other way around.

What does this have to do with AGT correspondence \cite{AGT}? As we will review, \cite{DVt} conjectured that  large $N$ duality of topological string provides a physical explanation for AGT correspondence, under certain conditions: Conformal block should admit free field representation, and Liouville theory should have central charge $c=1$ to correspond to topological string. 
 
 We interpret this purely in the gauge theory language, in the context of the gauge/vortex duality, and show that this leads to a proof of correspondence in a fairly general setting. The partition function of the 4d ${\cal N}=2$ gauge theory associated in \cite{G2, Gaiotto:2009hg} to a genus zero Riemann surface with arbitrary number of punctures equals the conformal block of Liouville theory with arbitrary central charge $c$, on the same surface. The free field representation of conformal blocks implies Coulomb moduli are quantized, but all other parameters remain arbitrary. The crucial role vortices play, extends AGT correspondence to a triality -- between the gauge theory, its vortices, and Liouville theory. The striking aspect of this result, which appeared first in \cite{AHKS}, is the simplicity of the proof. While in this review we focus on the simplest variant of AGT correspondence, relevant for Liouville theory, same ideas apply for more general Toda CFTs (Liouville theory corresponds to $A_1$ Toda). The generalization to $A_n$ Toda case can be found in \cite{AHS}.\footnote{Proofs of (some aspects of) AGT correspondence using different ideas appeared in \cite{Fateev:2009aw, Alba:2010qc, Mironov:2010pi, Morozov:2013rma, Braverman}. }

\section{Background}

Alday, Gaiotto and Tachikawa \cite{AGT} conjectured a correspondence between conformal blocks of Liouville CFT and partition functions of a class of four-dimensional theories, in 4d $\Omega$-background \cite{N2}. The 4d theories are conformal field theories with ${\cal N}=2$  supersymmetry defined in \cite{G2, Gaiotto:2009hg} (see also [V:1]) in terms of a pair of M5 branes wrapping a Riemann surface $C$, which we will call the Gaiotto curve. Specifying both the conformal block and the 4d theory ${\cal T}_{4d}$ in this class, involves a choice of the curve $C$ with punctures, data at the punctures and pants decomposition. The conjecture is often referred to as 4d/2d correspondence.

\subsection{4d Gauge Theory}

Let $\Sigma$ be the Seiberg-Witten curve of ${\cal T}_{4d}$,
\beq\label{4dcurve}
{ \Sigma}:\qquad \qquad p^{2} + \phi^{(2)}(z)=0.
\eeq
with meromorphic one form $\lambda = p dz$. ${\Sigma}$ is a double cover of $C$, $z$ is a local coordinate on $C$, and $\phi^{(2)}(z) (dz)^2$ is a degree 2 differential on $C$, whose choice specifies the IR data of the theory (the point on the Coulomb branch). Specifying the UV data of the theory requires fixing the behavior of the Seiberg-Witten differential $\lambda$ near the punctures.

At a puncture at $z=z_i$, the $\lambda$ has a pole of order $1$, with residues
$$
p\sim  \pm {\alpha_{i}\over z- z_i}
$$
on the two sheets. These lead to second order poles of $\phi^{(2)}(z)dz^2$.  In the gauge theory, $\alpha_i$'s and $z_j$'s are the UV data; the mass parameters and the gauge couplings. $\Sigma$ also depends on the IR data of the gauge theory, the choice of Coulomb branch moduli. These are associated to the sub-leading behavior of the ${\phi^{(2)}(z)}$ near the punctures.

Let
$${\cal Z}_{{\cal T}_{4d}}(\Sigma)
$$
be the partition function of the theory, in 4d $\Omega$-background.  Given a gauge theory description of ${\cal T}_{4d}$, ${\cal Z}_{{\cal T}_{4d}}(\Sigma)$ can be  computed using results of Nekrasov in \cite{N2} (see also [V:3]).
In addition to the geometric parameters entering $\Sigma$,
${\cal Z}_{{\cal T}_{4d}}$ depends on
$$\epsilon_1, \;\; \epsilon_2,
$$
 the two parameters of the $\Omega$ background \cite{N2}. ${\cal Z}$ can in principle depend on data beyond the geometry of $\Sigma$;  different choices of the pants decomposition can lead to different descriptions of the theory with different but related ${\cal Z}$'s.

\subsection{2d Liouville CFT}

The Liouville CFT has a representation in terms of a boson $\phi$:
$$ S_{Liouv.} = \int dz d{\bar z} \;\sqrt g \; [g^{z {\bar z}}\partial_z \phi \, \partial_{\bar z} \phi +  Q \phi R + e^{2 b \phi} ].
$$
Consider a conformal block on $C$ with insertions of primaries with momenta $\alpha_i$ at points $z_i$:
$$
{\cal B}(\alpha, z)=\langle V_{\alpha_0}(z_0) \cdots V_{\alpha_{\ell}}(z_{\ell}) V_{\alpha_{\infty}}(\infty)\rangle,
$$
where
$$
V_{\alpha}(z) = \exp\left( - \frac{\alpha}{b} \phi(z) \right)
$$
is the vertex operator of a primary with momentum $\alpha$.
Above, $Q$ is the background charge, $Q= b+{1\over b} $; Liouville theory with this background charge has central charge $c=1+6 Q^2$.
In addition to momenta and positions of the vertex operators inserted, the conformal block depends on the momenta in the intermediate channels; in denoting the conformal block by ${\cal B}( \alpha, z)$ we have suppressed the dependence on the latter.

\subsection{The correspondence}

The conjecture of \cite{AGT} is that the partition function ${\cal Z}_{{\cal T}_{4d}}(\Sigma)$ computes a conformal block of Liouville CFT on $C$:
$${\cal Z}_{{\cal T}_{4d}}(\Sigma)={\cal B}(\alpha, z),
$$
where $b$ is related to two parameters $\epsilon_{1,2}$ by
$$
b= \sqrt{{\epsilon_1 \over \epsilon_2}},
$$
while the parameters $\alpha_i$, $z_i$ of $\Sigma$ map to the corresponding parameters in the conformal block and the Coulomb branch parameters map to the momenta in intermediate channels.

\section{AGT and Large $N$ Duality}
In \cite{DVt} Dijkgraaf and Vafa explained the correspondence, in a particular case of the self-dual $\Omega$-background,
\beq\label{self}
\epsilon_1= g_s =- \epsilon_2,
\eeq
in terms of a large $N$ duality in topological string theory. The argument of \cite{DVt} has three parts, which we will now describe. As everywhere else in this review, we will focus on the case when the Gaiotto curve $C$ is genus zero. One can extend the argument more generally \cite{DVt}, as all the ingredients generalize to $\Sigma$ a double cover of an arbitrary genus $g$ Riemann surface $C$.

\subsection{The Physical and the Topological String }

The gauge theory partition function ${\cal Z}_{{\cal T}_{4d}}(\Sigma)$ in the self-dual $\Omega$-background is conjectured in \cite{DVt} to be the same as the partition function
$$
 Z(Y_{\Sigma})
$$
of the topological B-model on a Calabi-Yau manifold $Y_{\Sigma}$, with topological string coupling $g_s$. The Calabi-Yau $Y_{\Sigma}$ is a hyper surface
\beq\label{4dcy}
Y_{ \Sigma}:\qquad \qquad p^{2}  +\phi^{(2)}(z)=uv,
\eeq
with holomorphic three-zero form $du dp dz/u$. The geometry of $Y_{\Sigma}$ and the Seiberg-Witten curve $\Sigma$ \eqref{4dcurve} are closely related: the latter is recovered from the former by setting $u$ or $v$ to zero.

This is a consequence of two facts.  First, one observes that IIB string theory on $Y_{\Sigma}$ is dual to M-theory with an M5 brane wrapping $\Sigma$.\footnote{This follows by compactifying M-theory with M5 brane on $\Sigma$ on a $T^2$ transverse to the M5 brane. Since the $T^2$ is transverse to the branes, it does not change the low energy physics. By shrinking one of the cycles of the $T^2$ first, we go to down to IIA string with an NS5 brane wrapping $\Sigma$. T-dualizing on the remaining compact transverse circle, we obtain IIB on $Y_{\Sigma}$.} This gives us another way to obtain the same 4d, ${\cal N}=2$ theory ${\cal T}_{4d}$. Second, the partition function of IIB string theory on $Y_\Sigma$ times the self-dual $\Omega$ background is the same as the topological B-model string partition function on $Y_{\Sigma}$ \cite{N2, Losev:2003py, Hollowood:2003cv}. Thus, one can simply identify the physical and the topological string partition functions
\beq\label{gt}
{\cal Z}_{{\cal T}_{4d}}(\Sigma) =Z(Y_{\Sigma}).
\eeq
The power of this observation is that the topological B-model partition function is well defined even when the Nekrasov partition function is not -- because for example, the gauge theory lacks a Lagrangian description. It is also important that sometimes one and the same topological string background gives rise to several different Lagrangian descriptions for one and the same theory -- for example, $SU(2)^{l-2}$ with four fundamentals vs. $SU(l)$ with $2l$ fundamentals. The former is the theory which is usually associated in the AGT literature to Liouville theory on the sphere with $l+1$ punctures; the latter is the one that naturally comes out from our approach.

\subsection{Large $N$ Duality in Topological String}

 Next, \cite{DVt} show that the B-model on  $Y_\Sigma$ has a dual, holographic description, in terms of $N$ topological B-model branes on a different Calabi-Yau, related to $Y_{\Sigma}$, by a geometric transition. Let us first describe the Calabi-Yau that results. Then, we will explain the duality.
\subsubsection{A Geometric Transition}

By varying Coulomb branch moduli of ${\cal T}_{4d}$ we can get the Seiberg-Witten curve $\Sigma$ to degenerate. Let us call the degenerate curve that results the $S$-curve:
\beq\label{4dcurve2}
{S}:\qquad \qquad p^2 - (W'(z))^2 =0.
\eeq
Here
%
$$
W'(z)=\sum_{i=0}^{\ell} {\alpha_i\over z- z_i},
$$
is determined by keeping the behavior of the Seiberg-Witten differential fixed at the punctures. The $S$-curve describes the degeneration of the Seiberg-Witten curve to two components, $p\pm W'(z)=0$. Correspondingly, a single M5 brane wrapping $\Sigma$ breaks into two branes, wrapping the two components.

The $S$-curve corresponds to a singular Calabi-Yau $Y_S$:
\beq\label{4dcy2}
Y_{ S}:\qquad \qquad p^{2}- (W'(z))^2   =uv,
\eeq
with singularities at $u,v,p$ equal to zero and points in the $z$-plane where
$$
W'(z)=0.
$$
The Calabi-Yau we need is obtained by blowing up the singularities. One can picture this by viewing $Y_S$ as a family of $A_1$ surfaces, one for each point in the $z$-plane. At every $z$ there is an $S^2$ in the $A_1$ surface whose area is proportional to $|W'(z)|$, The singularity occurs where the $S^2$ shrinks. After blowing up, we get a family of $S^2$'s of non-zero area, one at each point in the $z$ plane, and all homologous to each other. The minimal area $S^2$'s are where the singularities were -- at points in the $z$ plane with $W'(z)=0$.

The geometric transition trades $Y_{\Sigma}$ for the blowup of $Y_{S}$. For economy of notations, we will denote $Y_S$ and its blowup in the same way, since their complex structure is the same, given by \eqref{4dcy2}.

\subsubsection{Large $N$ Duality}

The B-model on  $Y_\Sigma$ has a holographic description in terms of B-model on (the blowup of ) $Y_S$ with $N$ topological B-model D-branes wrapping the $S^2$ class.  The branes get distributed between the minimal $S^2$'s at points in the $z$-plane where $W'(z)$ vanishes. This breaks the gauge group from $U(N)$ to $\prod_{i=0}^{\ell} U(N_i)$, with $\sum_i N_i=N$. The Coulomb-branch moduli of $Y_{\Sigma}$ get related to t'Hooft couplings $N_i g_s$ in the theory on B-branes. The remaining parameters, $\alpha, z$ and the topological string coupling $g_s$ are the same on both sides. This is the topological B-string version of gauge/gravity duality \cite{DV}.

The large $N$ duality relates the closed topological string partition function of the B-model on $Y_\Sigma$, and thus the partition function ${\cal Z}(\Sigma)$, to partition function of the $N$ topological B-branes on (the blowup of) $Y_S$,
$$
{Z}(Y_\Sigma) = {Z}(Y_S; N).
$$
The right hand side depends not only on the net number of branes, but also how they are split between the different ${\mathbb P}^1$'s.

The partition function of $N$ B-type branes wrapping the $S^2$ in a Calabi-Yau of the form of \eqref{4dcy2} was found in \cite{DV}. It equals
\beq\label{matinx}
{1\over {\rm vol}(U(N))} \int d\Phi\; \exp( \,{\rm Tr}  W(\Phi)/g_s),
\eeq
where ${\rm vol}(U(N))$ is the volume of $U(N)$. The integral is a holomorphic integral, over $N\times N$ complex matrices $\Phi$. In evaluating it, one has to pick a contour, ending at a critical point of the potential.  In the present case,
$$
W(x) = \sum_i \alpha_i \log(x-z_i).
$$
Diagonalizing $\Phi$ and integrating over the angles, the integral reduces to
\beq\label{matin}
{Z}(Y_S; N)={1\over N!} \int d^N x \prod_{ I< J} (x_I-x_J)^2 \prod_{I, i}(x_I-z_i)^{\alpha_i/g_s}.
\eeq
Here $N!$ is the order of the Weyl group that remains as a group of gauge symmetries.

The claim is that large $N$ expansion of the integral equals topological B-model partition function on \eqref{4dcy}. At the level of planar diagrams this can be seen as follows. In the matrix integral, define an operator
\beq\label{mmphi}
\partial \phi(z) =   W'(z) +   g_s \sum_I {1\over z- x_I},
\eeq
where $x_I$ are the eigenvalues of $\Phi$.
The expectation value of
$$T(z) = ( \partial \phi)^2$$
computed in the matrix theory captures the geometry of the underlying Riemann surface by identifying $\phi^{(2)}(z)$ in \eqref{4dcurve} with
$$
\phi^{(2)}(z) =  \langle T(z) \rangle.
$$
There are two limits in which a classical geometry emerges from this. First, by simply sending $g_s$ to zero we recover the $S$-curve, since then $ \langle T(z) \rangle =(W')^2$. But, there is also a new classical geometry that emerges at large $N$. Letting $N_i$'s go to infinity, keeping $N_i g_s$ fixed we get
$$
\langle T(z) \rangle  \sim  (W'(z))^2+f(z),
$$
with
$$f(z) = \langle g_s  \sum_I {W'(z) - W'(x_I)\over z-x_I}\rangle.
$$
From the form of the potential $W(z)$, it follows that $f(z)$ has the form
$$
f(x) = \sum_i  {\mu_i \over x-z_i}
$$
with at most single poles. Thus, the branes deform the geometry of the Calabi-Yau we started with. The resulting Calabi-Yau is exactly of the form $Y_{\Sigma}$ \eqref{4dcy}, corresponding to the Seiberg-Witten curve $\Sigma$ in \eqref{4dcurve} at a generic point of its moduli space.

The large $N$ duality is expected to hold order by order in the $1/N$ expansion; we just gave evidence it holds in the planar limit (the full proof of the correspondence in the planar limit is easy to give along these lines, see \cite{DV}). The good variable in the large $N$ limit turns out to be the chiral operator $\phi(z)$ we defined in \eqref{mmphi}. The field $\phi(z)$, is in fact the string field of the B-model.

The B-model string field theory, called Kodaira-Spencer theory of gravity, was constructed in \cite{BCOV}, capturing variations of complex structure. For Calabi-Yau manifolds of the form \eqref{4dcy} the Kodaira-Spencer theory becomes a two dimensional theory on the curve $\Sigma$. The theory describes variations of complex structures of $Y_{\Sigma}$, so the Kodaira-Spencer field can be identified with fluctuations of the holomorphic $(3,0)$ form of the Calabi-Yau. For $Y_{\Sigma}$ fluctuations of the $(3,0)$ form are equivalent to fluctuations of the meromorphic $(1,0)$ form on $\Sigma$:
$$
 \delta \lambda = \delta p dz= \partial \phi dz.
$$
The Kodaira-Spencer field is a chiral boson $\phi$ which lives on $\Sigma$. When $\Sigma$ is a double cover of a curve $C$, a single boson on $\Sigma$ is really a pair of bosons $\phi_{1}$, $\phi_2$ on $C$, one corresponding to each sheet. The field $\phi$ that arises in the matrix model in \eqref{mmphi} can be thought of as off diagonal combination of the two. The diagonal combination is a center of mass degree of freedom and decouples from the dynamics of the branes.\footnote{The full topological string partition function in the presence of branes is given by the matrix integral in \eqref{matinx} - \eqref{matin}, describing open strings, times a purely closed topological string partition function of $Y_S$. This will be relevant later on.}
\subsection{Topological D-branes and Liouville Correlators}
To complete the argument, \cite{DVt} observe that the B-brane partition function ${Z}(Y_S; N)$ equals the Liouville correlator at $c=1$, when written in the free-field or Dotsenko-Fateev representation  \cite{Dotsenko:1984nm,Dotsenko:1984ad},
\beq\label{p2}
{Z}(Y_S; N) \;\;= \;\;{\cal B}(\alpha/g_s, z; N)|_{c=1}.
\eeq

One treats the Liouville potential as a perturbation and computes the correlator in the free boson CFT
{\fontsize{11pt}{0pt}
\beq\label{expect}
{\cal B}(\alpha, z; N) = \langle V_{\alpha_1}(z_1) \ldots V_{\alpha_\ell}(z_\ell) V_{\alpha_\infty}(\infty) \;\;\oint dx_1 S(x_1)\cdots \oint dx_N S(x_N)\rangle_0,
\eeq}
where we took the chiral half. Here, $S(z)$ is the screening charge
$$
S(z) = e^{2 b \phi(z)},
$$
whose insertions come from bringing down powers of the Liouville potential. It follows that
\eqref{expect} vanishes unless
$$
\frac{\alpha_{\infty}}{b} + \sum_{i=0}^\ell  \frac{\alpha_{i}}{b} = 2 b N + R Q,
$$
constraining the net $U(1)$ charge of the vertex operator insertions to be the number of screening charge integrals. This constraint can be found directly from the path integral, by integrating over the zero modes of the bosons \cite{Dotsenko:1984nm,Dotsenko:1984ad,Goulian:1990qr}. We will place a vertex operator at infinity of the $x$ plane, and then the equation determines the momentum of the operator at infinity in terms of the momenta of the $\ell+1$ remaining vertex operators at finite points and numbers of screening charge integrals.

An integral expression for the expectation value of the correlator in \eqref{expect} is easy to obtain, for example, by using the free boson mode expansion
$$
b \phi(z) = \phi_0 + h_0\,\mbox{log}\,z + \sum_{k\neq 0}h_k\frac{z^{-k}}{k},
$$
where $\phi_0$ is a constant, and $h_m$ satisfy the standard algebra
\beq
\label{heisenberg}
[h_{k}, h_{m}] = {-b^2 \over 2} \ k \,\delta_{k+m,0}
\eeq
where $k, m\in {\mathbb Z} $. From this one obtains the two point functions:
\begin{align*}
\nonumber \langle V_{\alpha}(z) V_{\alpha^{\prime}}(z^{\prime}) \rangle = (z - z^{\prime})^{\frac{-\alpha \alpha^{\prime}}{2 b^2}}, \\
\nonumber \langle V_{\alpha}(z) S(z^{\prime}) \rangle = (z - z^{\prime})^{\alpha}, \\
\langle S(z) S(z^{\prime}) \rangle = (z - z^{\prime})^{- 2 b^2}.
\end{align*}
The final result is that \eqref{expect} equals
$$
{\cal B}(\alpha, z; N)={r\over N!} \ \int \prod  d^Nx\;\prod_{i, I} (x_{ I} - z_i)^{\alpha_{ i}}
\;\prod_{J\leq I}(x_{I}-x_{J})^{-2 b^2},
$$
where the integrals are over the position of screening charge insertions and
$$
r = \prod\limits_{i, j} (z_i - z_j)^{\frac{-\alpha_i \alpha_j}{2 b^2}}
$$
is a constant, independent on the integration variables. This is the free-field $\beta$-ensemble (with $\beta = -b^2$) reviewed in [V:7].

Setting $\epsilon_1=-\epsilon_2$ (taking $b^2=-1$ in Liouville CFT) and rescaling $\alpha$ by $g_s$, it follows immediately that the free field expression for the conformal block ${\cal B}(\alpha /g_s, z; N)$ agrees with the partition function ${Z}(S; N)$ of B-branes in topological string on $Y_S$ as we claimed in \eqref{p2}. Moreover, in the large $N$ limit, the holomorphic part of the Liouville field $\phi(z)$ can be identified with the matrix model operator \eqref{mmphi}. This completes the argument of \cite{DVt}.

\subsection{Discussion}

The AGT conjecture, for $\epsilon_1+\epsilon_2=0$ can thus be understood as a consequence of a triality relating the closed B-model on $Y_{\Sigma}$, the holographic dual theory of $B$-branes on the resolution of $Y_S$ and the DF conformal blocks. The first two are conjectured to be related by large $N$ duality\footnote{It may be useful to summarize what the large $N$ asymptotic regime is, on each side of the correspondence. On the B-model side, it is sending $g_s$ to zero while keeping the combination $N g_s$ fixed. On the gauge theory side, it is sending $\epsilon_1 = -\epsilon_2$ to zero while keeping the Coulomb parameters fixed. On the Liouville side, it is sending all the momenta as well as the number $N$ of screening insertions to infinity, while keeping their ratios fixed.} in topological string theory, the latter two by the fact that the partition function of $B$-branes equals the DF block:
\beq\label{p2x}
\;\; Z(Y_{\Sigma}) \;\; {\stackrel{\rm{Large}\;N}{ = }}\;\;{Z}(Y_S; N) \;\;= \;\;{\cal B}(\alpha/g_s, z; N)|_{c=1}.
\eeq
We also used the embedding of topological string into superstring theory, which implies that the topological string partition function $Z(Y_{\Sigma})$ is the same as the physical partition function ${\cal Z}_{{\cal T}_{4d}}(\Sigma)$.

While this gives an explanation for the AGT correspondence in physical terms, it is by no means a proof: while the partition function of B-branes is manifestly equal to the Liouville conformal block in free field representation, the large $N$ duality is still a conjecture. The exact partition function of the B-model on $Y_{\Sigma}$ is not known, so one can only attempt a proof, order by order in the genus expansion. In addition, there is a string theory argument, but no proof, that the partition function of the gauge theory ${\cal Z}_{{\cal T}_{4d}}(\Sigma)$ and topological string partition function ${Z}(Y_{\Sigma})$ agree.  

Thirdly, from the perspective of the 4d gauge theory, it is very natural to consider the partition function on general $\Omega$-background, depending on  arbitrary $\epsilon_1$, $\epsilon_2$. Topological string on the other hand requires self-dual background, so the argument of \cite{DVt} can not be extended in this case.\footnote{For general $\epsilon_{1,2}$  the background does not simply decouple into a product of a Calabi-Yau manifold times the $\Omega$ background where the gauge theory lives. Turning on arbitrary $\Omega$ background requires the theory to have an $U(1)\in SU(2)_R$  R-symmetry to preserve supersymmetry. This requires the target Calabi-Yau manifold to admit a $U(1)$ action; this $U(1)$ action is used in constructing the background.} In \cite{DVt}, it was suggested to formulate the refinement at the level of  B-model string field theory. This remains to be developed better: refinement exists for any Calabi-Yau of the form $F(p,z)=uv$; the predictions from a naive implementation of this idea work for some, but not all choices of $F(p,z)$.

In the rest of the review, we will explain how to solve the last problem, and as it turns out the first two as problem as well, by following a different route.

The relation between topological string and superstring theory suggests one may be able to reformulate \cite{DVt} in string theory language, replacing topological string branes by branes in string or M-theory. While topological string captures the $\epsilon_1+\epsilon_2=0$ case only, the full superstring or M-theory partition function makes sense for any $\epsilon_1, \epsilon_2$. In fact, will will do something simpler yet: We will formulate the {\it gauge theory analogue of \cite{DVt}} for any $\epsilon_1$, $\epsilon_2$. We will see that this approach is powerful -- in fact it leads to a rigorous yet simple proof that the gauge theory partition function ${\cal Z}_{{\cal T}_{4d}}(\Sigma)$ agrees with the free field Liouville conformal block for $C$ a sphere with arbitrary number of punctures.

The triality of relations between the 4d gauge theory, its vortices, and Liouville conformal blocks which admit free field representation implies AGT correspondence, however it stops short of the most general case. The restriction to blocks that admit free field representation means, from the 4d perspective, that the Coulomb moduli are quantized to be  -- arbitrary -- integers, which get related to vortex charges on one hand, and numbers of screening charge integrals on the other. 

\section{Gauge/Vortex Duality}

Translated to gauge theory language, the large $N$ duality of topological string theory becomes a duality between the 4d ${\cal N}=2$ gauge theory ${\cal T}_{4d}$ and the 2d ${\cal N}=(2,2)$ theory on its vortices; we will denote the later theory ${\cal V}_{2d}$.  Observations of relations between the two theories go back to \cite{Dorey, DHT, HananyTong, HananyTong2}. Recently \cite{DH1, DH2} proposed that the two theories are dual -- indeed this is the "other" 2d/4d relation. On the face of it, the statement is strange at best:  to begin with, not even the dimensions of the 4d and the 2d theories match.

In this section we will show that, placed in a certain background, the 4d and the 2d theory describe the same physics, and thus there is good reason why their partition functions agree \cite{Simonstalk}. The large $N$ duality of \cite{DV, DVt} becomes a duality between two $d=2$, ${\cal N}=(2,2)$ theories: the 4d gauge theory ${\cal T}_{4d}$ we started with, in a variant of 2d $\Omega$-background with vortex flux turned on, and the 2d theory ${\cal V}_{2d}$ on its vortices.

\subsection{Higgs to Coulomb Phase Transition and Vortices}

In gauge theory language, the geometric transition that relates B-model on a Calabi-Yau $Y_{\Sigma}$, first to a singular Calabi-Yau $Y_S$ and then to a blowup of $Y_S$, is a Coulomb to Higgs phase transition.
This follows from embedding of the B-model into IIB superstring on a Calabi-Yau, and the relation between the string theory and the gauge theory which arises in its low energy limit \cite{Strominger:1995cz}. The same transition, in the language of M5 branes corresponds to degenerating a single M5 brane wrapping $\Sigma$, to a pair of M5 branes wrapping two Riemann surfaces $p\pm W'(z)=0$ that the $S$-curve consists of, and then separating these in the transverse directions (these are $x^{7,8,9}$ directions in the language of \cite{W7}).

The geometric transition becomes a topological string duality, as opposed to a phase transition, by adding $N$ B-branes on the $S^2$ in the blowup of $Y_S$. In terms of IIB string,  the $N$ B-branes on the $S^2$ are $N$ D3 branes wrapping the $S^2$ and filling 2 of the 4 space-time directions. In terms of M5 branes, the vortices are M2 branes stretching between the M5 brane wrapping $p-W'=0$ and the one wrapping $p+W'=0$. In the gauge theory on the Higgs branch, $N$ branes of string/M-theory become $N$ BPS vortices, as explained in  \cite{Greene:1996dh,Hori:1997zj} and \cite{HananyTong, HananyTong2}.\footnote{One should not confuse the vortices here with surface operators in the gauge theory, studied for example in \cite{WittenGukov, Alday:2009fs,Dimofte:2010tz}. The surface operators are solutions on the Coulomb branch, with infinite tension. From the M5 brane perspective, surface operators are semi-infinite M2 branes ending on M5's.}

The vortices in question are non-abelian generalization of Nielsen-Olesen vortex solutions  whose BPS tension is set by the value of the FI parameters. These were constructed explicitly in \cite{HananyTong, HananyTong2}. The net  BPS charge of the vortex is $N=\int {\rm Tr}F$ where $F$ is the field strength of the corresponding gauge group and the integral is taken in the 2 directions transverse to the vortex.\footnote{Usually, the gauge theories on M5 branes wrapping Riemann surfaces are said to be of special unitary type, rather than unitary type. There is no contradiction; the $U(1)$ centers of the gauge groups that arise on branes are typically massive by Green-Schwarz mechanism. This does not affect the BPS tension of the solutions, see e.g. discussion in \cite{Douglas:1996sw}.}

\subsection{Gauge/Vortex Duality}

Consider subjecting the 4d ${\cal N}=2$ theory ${\cal T}_{4d}$ to a {\it two}-dimensional ${\Omega}$-background in the two directions transverse to the vortex. We set $\epsilon_1=\hbar $ to zero momentarily since the duality we want to claim holds for any $\hbar$. This is the Nekrasov-Shatashvili background studied in \cite{NS2}.  The 2d $\Omega$-background depends on the one remaining parameter, $\epsilon=\epsilon_2$. (The equivalence of two theories is a stronger statement that the equivalence of their partition functions. The later assumes a specific background, while the former implies equivalence for any background. We will let $\hbar$ be arbitrary once we become interested in the partition functions, as opposed to the theories themselves.)

 As in \cite{NS2}, we view this partial ${\Omega}$-background as a kind of compactification: it results in a 2d theory with infinitely many massive modes, with masses spaced in multiples of $\epsilon$. The background also preserves only $4$ out of the $8$ supercharges.  Under conditions which we will spell out momentarily, the effective 2d ${\cal N}=(2,2)$ theory that we get is equivalent to the theory on its vortices. The condition that is clearly necessary is that we turn on vortex flux.  We assume it is also sufficient.

The vortex charge is $\int_D F_i =N_i$ where $i$ labels a $U(1)$ gauge field in the IR, and $F_i$ is the corresponding field strength. Here, $D$ is the cigar, the part of the 4d space time with 2d $\Omega$ deformation on it. It is parameterized by one complex coordinate, which we will call $w$. Without the $\Omega$ deformation,  turning on $N_i\neq 0$ would be introducing singularities in space-time which one would interpret in terms of surface operator insertions \cite{WittenGukov}. In $\Omega$ background, one can turn on the vortex flux without inserting additional operators -- in fact, the only effect of the flux is to shift the effective values of the Coulomb branch moduli. Let us explain this in some detail.

In the $\Omega$ background, $D$ gets rotated with rotation parameter $\epsilon$, in such a way that the origin is fixed. The best way to think of the theory that results \cite{NW, NS2} is in terms of deleting the fixed point of the rotation, and implementing a suitable boundary condition. Because the disk is non-compact, we really need two boundary conditions: one at the origin of the $w$ plane and one at infinity. Turning on flux simply changes the boundary condition we impose at the origin. Without vortices, one imposes the boundary condition \cite{NW} that involves setting $A_{i,w}=0$, where $A_{i,w}$ is the connection of $i$-th $U(1)$ gauge field along $D$. With $N_i$ units of vortex flux on $D$, we need instead $A_{i,w}= N_i/w $.

In the $\Omega$-background, the 4d theory in the presence of  $N_i$ units of vortex flux $A_{i,w}= N_i/w $ and with Coulomb branch scalar $a_i$ turned on is equivalent to studying the theory without vortices, at $A_{i,w}=0$, but with $a_i$ shifted by
$$
a_i \;\; \rightarrow \;\;  a_i+N_i \epsilon.
$$
This comes about because in the $\Omega$ background, $a_i$ always appears in the combination \cite{NW}

$$
a_i+\epsilon wD_{i,w},
$$
where $D_w=\partial_w + A_{i,w}$ is the covariant derivative along the $w$-plane traverse to the vortex.  Thus, in the $\Omega$ background, at the level of F-terms, turning on vortex flux is indistinguishable from the shift the effective values of the Coulomb branch moduli.\footnote{In \cite{NW} one proves that any flat gauge field on the punctured disk preserves supersymmetry of the $\Omega$ background.}

The 4d theory placed in 2d $\Omega$-background, with vortex flux turned on has an effective description studied in \cite{NS2, NW} in terms of the $2d$ theory with ${\cal N}=(2,2)$ supersymmetry with massive modes integrated out. The $(2,2)$ theory has a non-zero superpotential
${\cal W}(a, \epsilon; N) = {\cal W}_{NS}(a_i +N_i \epsilon, \epsilon)$,where ${\cal W}_{NS}(a_i, \epsilon)$ is the effective superpotential derived in \cite{NS2}, and the shift by $N_i \epsilon$ is due to the flux we turned on. The critical points of the superpotential correspond to supersymmetric vacua of the theory.  In the A-type quantization, considered in \cite{NS2},  the vacua are at $\exp(\partial_{a_i}  {\cal W}_{NS}/\epsilon) =1$ or, equivalently, at $a_{D,i}/\epsilon = \partial_{a_i} {\cal W}_{NS}/\epsilon \in {\mathbb Z}$. In the B-type quantization, they are at  $a_i/\epsilon \in {\mathbb Z}$ \cite{quantum, DH1,DH2}. Choosing $a_i=0$, for all $i$ is the vacuum at the intersection of the Higgs and the Coulomb branch. Choosing $a_i = N_i \epsilon$ corresponds to putting the theory at the root of the Higgs branch -- but in the background of $N_i$ units of flux.\footnote{We thank Cumrun Vafa for discussion relating to this point.}

There is a {\it second description} of the same system. If we place the theory at the root of the Higgs branch, the 4d theory has vortex solutions of charge $N_i$ even without the $\Omega$-deformation. These are the non-abelian Nielsen-Olsen vortices of \cite{HananyTong, HananyTong2}. We get a second $2d$ theory with ${\cal N}=(2,2)$ supersymmetry -- this is the theory on vortices themselves. In the theory on the vortex, the only effect of the $\Omega$-deformation is to give the scalar, parameterizing the position of the vortex in the $w$-plane, twisted mass $\epsilon$. From this perspective, turning on $\epsilon$ is necessary since it removes a flat direction (position of vortices in the trasverse space).

Similarity of the two theories at the level of the BPS spectrum was observed in \cite{Dorey, DHT, HananyTong, HananyTong2, Shifman:2004dr}. For a class of theories, this duality was first proposed in \cite{DH1, DH2}, motivated by study of integrability. The physical explanation for gauge/vortex duality we provided implies the duality should be general, and carry over to many other systems.\footnote{See \cite{AS3} for a highly nontrivial example.}

\subsection{Going up a Dimension}

The duality between ${\cal T}_{4d}$, in the variant of the 2d $\Omega$-background we described above, and ${\cal V}_{2d}$ lifts to a duality in one higher dimension, between a pair of theories, ${\cal T}_{5d}$ and ${\cal V}_{3d}$, compactified on a circle. We will prove the stronger, higher dimensional version, of the duality. ${\cal T}_{4d}$ lifts to a five-dimensional theory ${\cal T}_{5d}$ with ${\cal N}=1$ supersymmetry. From 4d perspective, one gets a theory with infinitely many Kaluza-Klein modes. One can view this theory as a deformation of ${\cal T}_{4d}$, depending on one parameter, the radius $R$ of the circle. Note that ${\cal T}_{5d}$ is not simply placed in a product of 2d $\Omega$-background times a circle -- rather the background is a circle fibration
$$( D\times S^1)_t,
$$
where as one goes around the $S^1$ D rotates by $t$, sending $w \rightarrow w t$.\footnote{This 3d background was used in \cite{N2, Losev:2003py, NO, NW} as a natural path to defining the 2d $\Omega$-background. For a review see \cite{Nekrasov:2004vw}.} Similarly, the 2d theory on the vortex, ${\cal V}_{2d}$ lifts to a 3d theory ${\cal V}_{3d}$, on a circle of the same radius.  The claim is that the two $d=2$, ${\cal N}=(2,2)$ theories we get in this way are dual, where the duality holds at least at the level of $F$-type terms. In the limit when $R$ goes to zero, the KK tower is removed, and we recover the theories we started with.

In the next section we will prove the duality by showing that partition functions of the two theories agree. When we compute the partition function of the 5d theory, we submit it to the full Nekrasov background depending on both $\epsilon$ and $\hbar$. This is the background
\beq\label{5dob}
(D \times {\mathbb C} \times S^1)_{q,t},
\eeq
where as one goes around the $S^1$, we simultaneously rotate $D$ by $t=e^{R \epsilon}$, and ${\mathbb C}$ by $q^{-1}=e^{- R \hbar}$. In the 3d theory on vortices, $\epsilon$ is a twisted mass, but $\hbar$ is a parameter of the $\Omega$ background along the vortex world volume. The background for ${\cal V}_{3d}$ is fixed once we choose the background for ${\cal T}_{5d}$, simply by the 5d origin of the vortices. ${\cal V}_{3d}$ is compactified on %
\beq\label{3dob}
({\mathbb C} \times S^1)_{q}.
\eeq
As we go around the $S^1$, ${\mathbb C}$ rotates by $q^{-1}$, and we turn on a Wilson line $t$ for a global symmetry rotating the adjoint scalar (and thus giving it mass $\epsilon$).

\section{Building up  Triality}

When ${\cal T}_{5d}$ is a lift of the M5 brane theory of section 2 to a one higher dimensional theory on a circle of radius $R$, the gauge/vortex duality extends to a triality. The triality is a correspondence between the 5d gauge theory
${\cal T}_{5d}$, the 3d theory on its vortices ${\cal V}_{3d}$, both on a circle of radius $R$ and a $q$-deformation of Liouville conformal block. As $R$ goes to zero, the $q$ deformation goes away and we recover the conformal blocks of Liouville.
The $q$-deformation of the Virasoro algebra was defined in \cite{Shiraishi:1995rp, Awata:1996xt}, and studied further and as well as extended to W-algebras in \cite{FR1}.

The triality comes about because the partition function of the vortex theory ${\cal V}_{3d}$ will turn out to equal the $q$-deformed Liouville conformal block,
\beq\label{first}
{\cal Z}_{{\cal V}_{3d}} ={\cal B}_{q},
\eeq
analogously to the way the partition function of topological D-branes was the same as the conformal block of Liouville at $b^2=-1$.
The relation between ${\cal T}_{5d}$ and ${\cal V}_{3d}$ is the gauge/vortex duality. The duality implies that their partition functions are equal,
\beq\label{second}
{\cal Z}_{{\cal T}_{5d}} = {\cal Z}_{{\cal V}_{3d}}.
\eeq
The left hand side is computed on \eqref{5dob} and the right hand side, by restriction, on \eqref{3dob}.  Thus, combining the two relations, we get a relation between $R$-deformation of the partition function of ${\cal T}_{4d}$ and the $q$-deformation of the Liouville conformal block,
\beq\label{main}
{\cal Z}_{{\cal T}_{5d}} = {\cal Z}_{{\cal V}_{3d}} = {\cal B}_{q}.
\eeq
In a limit, both deformations go away and we recover the relation between a partition function of the 4d, ${\cal N}=2$ theory ${\cal T}_{4d}$ and the ordinary Liouville conformal block ${\cal B}$.  We will prove this for the case when $C$ is a sphere with any number of punctures. The equality in \eqref{second}, as we anticipated on physical grounds, holds for special values of Coulomb branch moduli -- those corresponding to placing the 5d theory at a point where the Higgs branch and Coulomb branches meet, and turning on fluxes. By taking the large flux limit, where $N_i$ goes to infinity, $\epsilon$ goes to zero keeping their product $N_i\epsilon$ fixed, all points of the Coulomb branch and arbitrary conformal blocks get probed in this way.

In the rest of the section we will spell out the details of the theories involved, and their partition functions. Then, in the next section, we will prove their equivalence.

\subsection{The 5d Gauge Theory ${\cal T}_{5d}$}

The 5d ${\cal N}=1$ theory ${\cal T}_{5d}$ per definition reduces to, as we send $R$  to zero, the 4d theory ${\cal T}_{4d}$ arising from a pair of M5 branes wrapping a genus zero curve $C$ with $\ell+2 $ punctures.

The ${\cal T}_{5d}$ theory turns out to be very simple: at low energies it is described by a $U(\ell)$ gauge theory with $2\ell$ hyper-multiplets: $\ell$ hypermultiplets in fundamental representation, $\ell$ in anti-fundamental, and 5d Chern-Simons level zero.\footnote{At very short distances there is a UV fixed point corresponding to it, which is a strongly coupled theory, accessible via its string or M-theory embedding \cite{Seiberg:1996bd,Intriligator:1997pq}} Except for $\ell =2$, the $U(\ell)$ gauge theory theory is different from the generalized quiver of \cite{G2}. This is nothing exotic: there are different ways to take $R$ to zero limit, and different limits can indeed result in inequivalent theories. At finite $R$, the theory we get is unique, but with possibly more than one description.

The Coulomb branch of the 4d theory ${\cal T}_{4d}$ is described by a single M5 brane wrapping the 4d Seiberg-Witten curve \eqref{4dcurve}. The Seiberg-Witten curve of ${\cal T}_{5d}$ compactified on a circle can be written as
\beq\label{5dsw}
\Sigma:\qquad Q_+(e^x) e^p +P(e^x) + Q_-(e^x) e^{-p} =0,
\eeq
with the meromorphic one form equal to $\lambda = p dx$ (see, e.g. \cite{Nekrasov:1996cz}). We will denote both the 4d and the 5d Seiberg Witten curves by the same letter, $\Sigma$ even though the curves are inequivalent; it should be clear from the context which one is meant. Here, $Q_{\pm}$ are polynomials of degree $\ell$ in $e^x$,
$$
Q_{\pm}(e^x) =  {e^{\pm \zeta/2 } \prod_{i=1}^{\ell} ( 1- e^x/  f_{\pm, i}) },
$$
and $P(x)$ is a polynomial of degree $\ell$ in $x$. At points where the Higgs and the Coulomb branch meet, $\Sigma$ degenerates to:
\beq\label{5dcurve2}
S\;\;:\qquad\qquad
 (Q_+(e^x) e^{p} - Q_-(e^x) )  (e^{-p} - 1)=0.
\eeq
The 5d Seiberg-Witten curve in \eqref{5dsw} and the S-curve in \eqref{5dcurve2} reduce to the 4d ones in \eqref{4dcurve}, and \eqref{4dcurve2}, by taking the $R$ to zero limit. The limit one needs corresponds to keeping $\zeta/R$ and  $p/R$ fixed and taking
\beq\label{massb}
f_{+, i} = z_i, \ \ \ f_{-,i} = z_i \ q^{\alpha_i}.
\eeq
Finally, one defines $z=e^x$, and replaces $p$ by $pz$ to get \eqref{4dcurve2}, the curve with its canonical one form $\lambda = pdz$. Note that one of the punctures we get is automatically placed at $z=0$.\footnote{The second four-dimensional limit gives the 4d ${\cal N}=2$ $U(\ell)$ gauge theory with $2\ell$ fundamental hypermultiplets by \cite{W7, G2}. In the Seiberg-Witten curve, one writes $f_{i}$ as $f_i = e^{R \mu_i}$, and takes $R$ to zero keeping $x/R$, $e^p R$, $e^{\zeta}R$ and the $\mu$'s fixed in the limit. The effect of this is that the 4d curve has the same form as \eqref{5dsw}, but with $Q$ and $P$ replaced by polynomials of the same degree, but in $x$, rather than $e^x$.}

\subsubsection{Partition function in $\Omega$-background}

The 5d $\Omega$-background is defined as a twisted product
\beq\label{back}
({\mathbb C}\times {\mathbb C}\times S^1)_{q,t},
\eeq
where as, one goes around the $S^1$, one rotates the two complex planes by $q = \exp(R \epsilon_1)$ and $t^{-1}=\exp(R \epsilon_2)$ (the first copy of ${\mathbb C}$ is what we called $D$ before). These are paired together with the 5d $U(1)_R\subset SU(2)_R$ symmetry twist by $t q^{-1}$, to preserve supersymmetry. The 5d gauge theory partition function in this background is the trace
\beq\label{5dtrace}
{\cal Z}_{{\cal T}_{5d}}(\Sigma)={\rm Tr} (-1)^F {\bf g}_{5d},
\eeq
corresponding to looping around the circle in \eqref{back}. Insertion of $(-1)^F$ turns the partition function of the theory to a supersymmetric partition function. One imposes periodic identifications with a twist by ${\bf g}$ where ${\bf g}$ is a product of simultaneous rotations: the space-time rotations by $q$ and $t^{-1}$, the $R$-symmetry twist,  flavor symmetry rotations $f_{i, \pm} = \exp(-R m_{i, \pm})$, and gauge rotation by $e_i = \exp(R a_i)$ for the $i$'th $U(1)$ factor. The latter has the same effect as turning on a Coulomb-branch modulus $a_i$ (see \cite{Nekrasov:2004vw} for a review). The partition function of ${\cal T}_{5d}$ in this background is computed in \cite{N2}, using localization. The partition function is a sum
\beq\label{bN}
{\cal Z}_{{\cal T}_{5d}}(\Sigma) = r_{5d} \sum_{\vec R} I^{5d}_{\vec R},
\eeq
over $\ell$-touples of 2d partitions
$$
{\vec R} = (R_1, \ldots , R_\ell),
$$
labeling fixed points in the instanton moduli space. The instanton charge is the net number of boxes $|\vec R|$ in the $R$'s.  The coefficient $r_{5d}$ contains the perturbative and the one loop contribution to the partition function.

The contribution
$$I^{5d}_{\vec R} = \;q^{\zeta |\vec R|} \ z_{V, {\vec R}} \times z_{H, {\vec R}} \times z_{H^{\dagger}, {\vec R}}
$$
of each fixed point is a product over the contributions of  the $U(\ell)$ vector multiplets, the $\ell$ fundamental and anti-fundamental hypermultiplets $H$, $H^{\dagger}$ in ${\cal T}_{5d}$. The instanton counting parameter, related to the gauge coupling of the theory, is $q^{\zeta}$. $I^{5d}$ depends on $\ell$ Coulomb branch moduli encoded in ${\vec e}$, and the $2\ell$ parameters ${\vec f}$ related to the masses of the $2\ell$ hypermultiplets.
The vector multiplet contributes
$$
z_{V, {\vec R}}= \prod_{1\leq a,b\leq \ell}[N_{R_a R_b}(e_a/e_b)]^{-1}.
$$
The $\ell$ fundamental hypermultiplets contribute
$$
z_{H, {\vec R}} = \prod_{1\leq a \leq \ell} \prod_{1\leq b \leq \ell}N_{\varnothing R_b}( v f_{a}/e_b),
$$
and the $\ell$ anti-fundamentals give
$$
z_{H^{\dagger}, {\vec R}} = \prod_{1\leq a \leq \ell} \prod_{1\leq b \leq \ell}N_{R_a \varnothing }( v e_a/{f_{b+\ell}}).
$$
The basic building block is the Nekrasov function
\begin{align*}
N_{RP}(Q) = \prod\limits_{i = 1}^{\infty} \prod\limits_{j = 1}^{\infty}
\dfrac{\varphi\big( Q q^{R_i-P_j} t^{j - i + 1} \big)}{\varphi\big( Q q^{R_i-P_j} t^{j - i} \big)} \
\dfrac{\varphi\big( Q t^{j - i} \big)}{\varphi\big( Q t^{j - i + 1} \big)},
\end{align*}
with $\varphi(x) = \prod\limits_{n=0}^{\infty}(1-q^n x)$ being the quantum dilogarithm \cite{Faddeev:1993rs, AHKS}. Furthermore,
$ T_R =(-1)^{|R|} q^{\Arrowvert R\Arrowvert/2}t^{-\Arrowvert R^t\Arrowvert/2}$, and $v = {(q/t)^{1/2}}$ as before (we use the conventions of \cite{Awata:2008ed}). In what follows, it is good to keep in mind that there is no essential distinction between the fundamental and anti-fundamental hypermultiplets.\footnote{By varying the Coulomb branch and the mass parameters, the real mass $m$ of the 5d hypermultiplet can go through zero. This exchanges the fundamental hypermultiplet of mass $m$ for an anti-fundamental of mass $-m$, while at the same time the 5d Chern-Simons level jumps by $1$ \cite{Witten5dphases}. A relation between the anti-fundamental and the fundamental hypermultiplet contributions to the partition function reflects this, see \cite{AHKS} for details.}
 In keeping with this, it is natural to think of all the $2\ell$ matter multiplets at the same footing,
and write the partition function, say, in terms of the fundamentals alone, whose masses run over $2\ell$ values, $f_a, f_{\ell+a}$, with $a=1, \ldots, \ell$.

\subsection{The Vortex Theory ${\cal V}_{3d}$}

The non-abelian generalization of Nielsen-Olesen vortices was found in \cite{HananyTong, HananyTong2}. In particular, starting with a bulk non-abelian gauge theory like ${\cal T}_{5d}$, with $8$ supercharges, $U(\ell)$ gauge symmetry and $2\ell$ hypermultiplets in fundamental representation,  they constructed the theories living on its half BPS vortex solutions. The theory on charge $N$ vortices is very simple: it is a $U(N)$ gauge theory with $4$ supercharges, with $\ell$ chiral multiplets in fundamental, and $\ell$ in anti-fundamental representation, as well as a chiral multiplet in the adjoint representation. The theory has a $U(\ell)\times U(\ell)$ flavor symmetry rotating the chiral and anti-chiral multiplets separately. This symmetry prevents their superpotential couplings. Since ${\cal T}_{5d}$ is five dimensional, the theory on its vortices is three dimensional ${\cal N}=2$ theory, which we will denote ${\cal V}_{3d}$. Presence of the 2d $\Omega$ background transverse to the vortex gives the adjoint chiral field twisted mass $\epsilon$. In addition, the theory is compactified on a circle of radius $R$. The masses of $2\ell$ hypermultiplets of ${\cal T}_{5d}$ get related to the $2\ell$ twisted masses of the chiral multiplets in ${\cal V}_{3d}$. We will see the precise relation momentarily.

\subsubsection{Partition function in $\Omega$-background}

We compactify ${\cal V}_{3d}$ on the 3d $\Omega$ background:
$$
({\mathbb C} \times S^1)_q.
$$
As we go around the $S^1$ we simultaneously rotate the complex plane by $q$ and twist by the $U(1)_R$-symmetry, to preserve supersymmetry. The partition function of the theory in this background in computes the index
\beq\label{3dtrace}
{\cal Z}_{{\cal V}_{3d}}(S; N)={\rm Tr} (-1)^F{\bf g}_{3d},
\eeq
where ${\bf g}_{3d}$ is a product of space-time rotation by $q$, an $U(1)_R$ symmetry transformation by $q^{-1}$, as well as the global symmetry rotation by $t$. The partition function of the theory can be computed by first viewing the $U(N)$ symmetry as a global symmetry: in this case, since the theory is not gauged, and due to the 3d $\Omega$ background, the index in \eqref{3dtrace} is simply a product of contributions from matter fields and the $W$-bosons, all depending on the $N$ Coulomb branch parameters $x_I$.

The contribution of the flavor in the fundamental representation is
\beq\label{basic}
\Phi_{F}(x)= \prod_{1\leq I \leq N}  {\varphi(e^{R x_I - R m_-})\over \varphi(e^{R x_I - R m_+})},
\eeq
where $m_{\pm}$ are the twisted masses. The right hand side is written in terms of Faddeev-Kashaev quantum dilogarithms \cite{Faddeev:1993rs, AHKS},
$$
\varphi(z) = \prod_{n=0}^{\infty}( 1 -q^n z).
$$
There are different ways to show this, for example, one can reduce the 3d theory down to quantum mechanics on the circle and integrate out a tower of massive states. Alternatively, the index can be obtained by counting holomorphic functions on the target space of the quantum mechanics, see \cite{Nekrasov:2004vw}. We can think of the flavor in the fundamental representation in one of two equivalent ways: it is a pair of ${\cal N}=2$ chiral multiplets, one in the fundamental and the other in the anti-fundamental representation. Alternatively, it contains a chiral multiplet and an anti-chiral multiplet, but both transform in the fundamental representation. The above way of writing $\Phi_F(x)$ is adapted to the second viewpoint.

The ${\cal N}=4$ vector multiplet, the adjoint chiral field and the $W$-bosons, give a universal contribution for any $U(N)$ gauge group:%
\beq
\Phi_{V}(x) = \prod_{1\leq I <J \leq N}{\varphi(\; e^{R x_I - R x_J})\over \varphi(t\; e^{R x_I - R x_J})}.
\eeq
The numerator is due to the W-bosons, and the denominator to the adjoint of mass scalar of mass $\epsilon$.
Finally, since the gauge group is gauged, we integrate over $x$'s. This simply projects to gauge invariant functions of the moduli space,
\beq\label{part}
{\cal Z}_{{\cal V}_{3d}}(S; N)={1\over N!}\int d^Nx \;\;  \Phi_{V}(x)\,  \prod_{a=1}^{\ell} \Phi_{F_a}(x)  \; e^{ \zeta \,{\rm Tr} x/\hbar}.
\eeq
The integrand is a product including all contributions of the massive BPS particles in the theory, the $W$ bosons, flavors $\Phi$'s, and the adjoint.
The exponent  contains the classical terms, the FI parameter $\zeta$, and the Chern-Simons level $k$ which is zero in our case. If the gauge symmetry were just a global symmetry, $x$'s would have been parameters of the theory and the partition function of the theory would have been the integrand. Gauging the $U(N)$ symmetry corresponds to simply integrating\footnote{This partition function is the index studied in \cite{AS, AS2, Aganagic:2012au} with application to knot theory; see also \cite{Fuji:2012nx}. The index is a chiral building block of the $S^3$ or $S^2\times S^1$ partition functions \cite{Hama:2010av,Kapustin:2011jm, Hama:2011ea,Pasquetti:2011fj, Nieri:2013yra, BDP, Taki:2013opa}, deformed by $t$, the fugacity of a very particular flavor symmetry. } over $x$.

We need to determine the contour of integration to fully specify the path integral.
The choice of a contour in the matrix model corresponds to the choice of boundary conditions at infinity in the space where the gauge theory lives \cite{Cheng:2010yw}. At infinity, fields have to approach a vacuum of the theory. For small $q$ and $t$, the vacua are the critical points of
$$
W(x) = \sum\limits_{a=1}^{\ell} \ \log {\varphi(e^{R x - R m_{-,a}})\over \varphi(e^{R x - R m_{+,a}})}.
$$
There are $\ell$ vacua of $W(x)$ both before and after the $R$-deformation. Splitting the $N$ eigenvalues so that $N_a$ of them approach the $a$-th critical point, we break the gauge group,
$$
U(N) \qquad \rightarrow \qquad U(N_1)\times \ldots \times U(N_\ell).
$$
We can think of all the quantities appearing in the potential as real; then the integration is along the real $x$ axis. To fully specify the contour of integration, we need to prescribe how we go around the poles in the integrand. The integral can be computed by residues, with slightly different prescriptions for how we go around the poles for the different gauge groups. In this way, we get $\ell$ distinct contours
${\cal C}_{N_1, \ldots, N_\ell}$, and with them the partition function,
$$
{\cal Z}_{{\cal V}_{3d}}(S;{N})={1\over \prod_{a=1}^{\ell} N_a!}\oint_{{\cal C}_{N_1, \ldots, N_\ell}} d^Nx \;\Phi_{V}(x)\\;
\prod_{a=1}^{{ \ell}} \Phi_{F_a}(x)   \; e^{-\zeta \,{\rm Tr} x/\hbar}.
$$
Dividing by $N_a!$ corresponds to dividing by the residual gauge symmetry, permuting the $N_a$ eigenvalues in each of the vacua. For $q=t$ this is a topological string partition function of the B-model on $Y_S$ studied in \cite{Aganagic:2002wv}, and related to Chern-Simons theory. The $q\neq t$ partition function is the partition function of refined Chern-Simons theory \cite{AS}, with observables inserted.

We will show that the partition function of ${\cal V}_{3d}$ is nothing but the $q$-deformation of the free-fieldfree field conformal block of the Liouville CFT on a sphere with $\ell+2$ punctures. Since the $q$ deformation of Liouville CFT might be not familiar, let us review it.

\subsection{$q$-Liouville}

In this section, we will show that the free field integrals of a $q$-deformed Liouville conformal field theory \cite{Shiraishi:1995rp, Awata:1996xt, Awata:2010yy} have a physical interpretation.  They are partition functions of the 3d ${\cal N}=2$ gauge theory, which we will called ${\cal V}_{3d}$, in the 3d $\Omega$-background $({\mathbb C }\times S^1)_q$. The equivalence of the $q$-Liouville conformal block and the gauge theory partition function is manifest. The screening charge integrals of DF are the integrals over the Coulomb branch of the gauge theory. Inserting the Liouville vertex operators corresponds to coupling the 3d gauge theory to a flavor. The momentum and position of the puncture are given by the real masses of the two chirals within the flavor.

The $q$-deformed Virasoro algebra is written in terms of the deformed screening charges
$$
S(z)=\ : \exp\left( 2 \phi_0 + 2 h_0 \log z + \sum\limits_{k \neq 0} \dfrac{1 + (t/q)^k}{k} h_k z^{-k} \right) :,
$$
where
\begin{align*}
[h_k, h_m] = \dfrac{1}{1 + (t/q)^k} \frac{1 - t^k}{1 - q^k} \ m\,  \delta_{k+m,0}.
\end{align*}
The defining property of the generators of the $q-$deformed Virasoro-algebra, is that they commute with the integrals of the screening charges $S$.
The primary vertex operators get deformed as well. The vertex operator carrying momentum $\alpha$ becomes:
$$
 V_{\alpha}(z) = \ : \exp\left( - \frac{\alpha}{b^2} \phi_0 - \frac{\alpha}{b^2} h_0 \log z + \sum\limits_{k \neq 0} \dfrac{1 - q^{-\alpha k}}{k(1 - t^{-k})} h_k z^{-k} \right) : \\.
$$
Note, that these operators manifestly become the usual Liouville operators in the limit where $q=e^{R\epsilon_1}, t=e^{-R \epsilon_2}$ go to $1$, by sending $R$ to zero.

Just as before, using these commutation relations, one computes the correlator and obtains the following free field integral:
\begin{align}\label{lcv}
{\cal B}_q(\alpha, z; N)=  {{r} \over \prod_{a=1}^{\ell}  N_a!} \oint_{{\cal C}_1, \ldots, {\cal C}_{\ell}} d^{N}y\; \Delta^2_{q,t}(y) \; \prod_{a=0}^{\ell} V_a(y; z_a),
\end{align}
where the measure is the $q,t$-deformed Vandermonde
$$
\Delta_{q,t}^2(y) = \prod_{1\leq I\neq J\leq N} {\varphi(y_I/y_J)\over \varphi(t\; y_I /y_J)},
$$
and the potential equals
$$
V_a(y; z_a) =  \prod\limits_{I=1}^{N} \dfrac{ \varphi\big(q^{\alpha_a} {z_a/y_I}\big) }{ \varphi\big(z_a/y_I\big) }.
$$
In particular, using the properties of the quantum dilogarithm, it is easy to find that
$V_0( y; 0) = (y_1 \ldots y_N)^{\alpha_0}.
$
As in the undeformed case, the relation holds up to a constant of proportionality $r$. In this paper, we avoid detailed consideration of this normalization constant. The meaning of the constant $r$, on the Liouville side, is to account for all possible two-point functions between the vertex operators $V_{\alpha}(z_a)$. Like in the undeformed case, the $N$ eigenvalues are grouped into sets of size $N_a$, $a=1,\ldots, \ell$, by the choice of contours they get integrated over.\footnote{The contours of integration \emph{are the same as} in the undeformed case -- encircling the segments $[0, z_a]$. The $q$ deformation affects the operators and the algebra, but not the contours.  It is important to emphasize that these contours agree with the alternative approach \cite{Mironov:2011dk} where the free field integrals are replaced by Jackson $q$-integrals: in our picture, the latter are the residue sums for the former.}

\section{Gauge/Liouville Triality}

In what follows, we will prove that there is a triality that relates the 5d and 3d gauge theories ${\cal T}_{5d}$ and ${\cal V}_{3d}$, compactified on a circle, and $q$-deformation of Liouville conformal blocks. We will show this in two steps.

\subsection{$q$-Liouville and ${\cal V}_{3d}$ }
The first step is to show that $q$-deformation of the Liouville conformal block \eqref{lcv}, corresponding to a sphere with $\ell+2$ punctures equals the partition function of ${\cal V}_{3d}$:
$$
{ {\cal Z}}_{{\cal V}_{3d}}(S; N)= { {\cal B}_q}(\alpha, z; N).
$$
This follows immediately by a simple change of variables that sets
\beq\label{chv}
z_a=e^{- R m_{+,a}}, \;\;q^{\alpha_a}=e^{R m_{+,a} - R m_{-,a}}, \;\;y = e^{- R x}.
\eeq
The insertion of a primary vertex operator in Liouville gets related to coupling the 3d gauge theory on the vortex to a flavor: the  mass splitting is related to Liouville momentum, the mass itself to the position of the vertex operator. The puncture at $z=0$ arises from the Fayet-Iliopolous potential, if we set  $\alpha_0=\zeta/\hbar-1$.
%

\subsection{${\cal V}_{3d}$ and ${\cal T}_{5d}$: Gauge/Vortex Duality }
The second step is to show that the partition function of the 5d gauge theory ${\cal T}_{5d}$ and partition function of its vortices, described by the 3d gauge theory ${\cal V}_{3d}$ agree
$$
{{\cal Z}}_{{\cal V}_{3d}}(S, N) =  {{\cal Z}}_{{\cal T}_{5d}}({\Sigma}).
$$
For this we place ${\cal T}_{5d}$ at the point where the Coulomb and Higgs branches of ${\cal T}_{5d}$ meet, $e_a =  f_{a}\, /v$ with $v = {(q/t)^{1/2}}$ as before, and $\Sigma$ degenerates to $S$. In addition we turn on $N_a$ units of vortex flux.\footnote{The shift by $v$ is due to the $\Omega$ background. It is natural that the partition function becomes singular at the point where the two branches meet; this determines the shift.} In the $\Omega$-background this is equivalent to not turning on flux and shifting the Coulomb-branch parameters of ${\cal T}_{5d}$ so that
$$
{\cal Z}_{{\cal T}_{5d}}(\Sigma) = r_{5d} \sum_{\vec R} I^{5d}_{\vec R}
$$
is evaluated at
\beq\label{rhc}
e_a =  t^{N_a}\,f_{a}\, /v,
\eeq
where $a$ runs form $1$ to $\ell$. Here, $f_a$ are the masses of $\ell$ of the $2\ell$ hypermultiplets, and the integer shifts correspond to $N_a$ units of vortex flux turned on. Note that as long as $N_a$ are arbitrary, this is no restriction at all.

To recover ${\cal T}_{5d}$ at an arbitrary point of its Coulomb branch, we take the limit $N_a\rightarrow \infty$, $\epsilon = {\rm ln}( t) \rightarrow 0$ keeping the product $N_a\epsilon$ fixed. The gauge/vortex duality is the gauge theory realization of large $N$ duality.

\subsubsection{Residues and Instantons}
We start by computing the partition function of ${\cal V}_{3d}$ by residues. Then we show that the sum over the residues is the instanton sum of the 5d gauge theory ${\cal T}_{5d}$. The positions of the poles are labeled by tuples of partitions, and the integrands are equal to Nekrasov summands.

With the change of variables in \eqref{chv}, the 3d partition function of ${\cal V}_{3d}$ becomes:
\beq\label{liouville}
{\cal Z}_{{\cal V}_{3d}}(N; S) ={1\over \prod_{a=1}^\ell N_a! }\oint_{{\cal C}_{1}, \ldots {\cal C}_\ell} d^{N} y\;\; I^{3d}(y),
\eeq
where the integrand $I^{3d}(y)$ equals
$$
I^{3d}(y) = V_0(y)\; \Phi_V(y)\;
\prod_{a=1}^{{ \ell}} \Phi_{F_a}(y),
$$
and, in terms of the new variables,
$$
 {\Phi}_{V}(y)=  \prod_{1\leq I \neq J\leq N}  {\varphi(y_J/y_I)\over\varphi(t y_J/y_I  )}, \;\;
 {\Phi}_{F_a}(y)=  \prod_{I=1}^{N}  {\varphi(q^{\alpha_a}  z_a/y_I)\over\varphi(z_a/y_I  )}, \;\; V_0(y) = \prod_{I=1}^N y_I^{\alpha_0}.
$$
The $\ell$ contours ${{\cal C}_{1}, \ldots {\cal C}_\ell} $  run around the intervals in the complex $y$ plane:  ${\cal C}_a$ circles the interval from $y=0$ to $y=z_a $, where $z_a$ is the location of a pole in the integral corresponding to a chiral multiplet going massless. The quantum dilogarithm  $\varphi( y) =  \prod_{n=0}^{\infty} {( 1- q^n \,y)}$ \cite{Faddeev:1993rs, AHKS} has zeros at $y=q^{-n}$, hence the integrand has poles there. The contour is chosen so as to pick up the residues of the poles. For each of the $\ell$ the groups of eigenvalues we choose the contour that runs from $0$ to $z_a$, circling the poles at
$$ y=  q^n\,z_a, \qquad n=0, 1,\ldots .
$$
For $|t|, |q|<1$,  the poles interpolate between $y=0$ and $y= z_a$, and the contours ${\cal C}_a$ circle around the interval (this is also where the critical points of the integral are located). However, not all the poles contribute -- the numerator in  ${\Phi}_{V}(y)$ eliminates some: all those for which poles for a pair $y_I, y_J$ coincide up to a $q$ shift. At the same time, the denominator of ${\Phi}_V(y)$ introduces new poles with $y$'s shifted by $t$, up to a multiple of $q$. Up to permutations, the poles that end up contributing are labeled by $\ell${\it-tuples of 2d Young diagrams}:
\beq\label{bmp}
{\vec R} = ({R}_1, \ldots, {R}_a, \ldots, { R}_\ell),
\eeq
where $R_a$ has at most $N_a$ rows. The poles corresponding to the $a$-th group of variables are at
$$
{y} = {y}_{\vec R},
$$
where, up to permutations the components of ${ y}_{\vec R}$ equal
\beq\label{cmp}
y_{(N_1+\ldots +N_{a-1})+ i} = q^{R_{a,i}} t^{N_a - i}z_a,
\eeq
where $i$ runs from $1$ to $N_a$ and $a$ from $1$ to $\ell$. The sum over the residues of the integral becomes the sum over the Young diagrams
$$
\prod_{a=1}^\ell {1\over N_a! }\;\oint_{{\cal C}_{1}, \ldots {\cal C}_\ell} d^{N} y\qquad \rightarrow \qquad \sum_{{\vec R}}.
$$
While the integrand itself does not make sense at a pole, the ratio of its values at different poles turns out to be finite. This implies that {\it ratio of the residues} at the poles labeled by ${\vec R}$ and ${\vec \varnothing}$
$$
I^{3d}_{{\vec R}} = {\rm res}^{-1}_{\varnothing}\cdot {\rm res}_{R} \; I^{3d}(y)
$$
is simply equal to the {\it ratio of the integrand} itself at the two poles:
\beq\label{3dsummand}
I^{3d}_{{\vec R}}= q^{\alpha_0|{\vec R}|}\cdot {\Phi_V(y_{\vec R})\over \Phi_V(y_{\vec {\varnothing}})}\cdot{
\prod_{a=1}^{{ \ell}} \Phi_{F_a}(y_{\vec R}) \over \;
\prod_{a=1}^{{\ell}} \Phi_{F_a}(y_{\vec {\varnothing}})}.
\eeq
Note that
${V_0(y_{\vec R})\over V_0(y_{\vec {\varnothing}})}= q^{\alpha_0|{\vec R}|} .
$
This makes the sum over residues easy to find:
$$
{\cal Z}_{{\cal V}_{3d}}(N; S) =r_{3d} \sum_{{\vec R}}  \; I^{3d}_{\vec R}(N, f),
$$
where
$$r_{3d} = {\rm res}_{\varnothing} I^{3d}(y).
$$
The structure of the answer is reminiscent of the 5d partition function ${\cal Z}_{{\cal T}_{5d}}(\Sigma)$, except that the sum in
${\cal Z}_{{\cal T}_{5d}}(\Sigma)$ runs over $\ell$-touples of Young diagrams of arbitrary size.

However, from the gauge/vortex duality, we only expect the 3d and the 5d partition functions to equal on the locus \eqref{rhc}.
Restricting to the locus \eqref{rhc}, the Nekrasov sum truncates to a sum over diagrams $R_a$ with at most $N_a$ rows. Moreover, for every such $\ell$-touple, the summand $I^{5d}_{{\vec R}}$ indeed becomes equal to $I^{3d}_{\vec R}$. The detailed proof is presented in \cite{AHKS}, here we only give a sketch.

Recall
$$
I^{5d}_{\vec R}=  q^{\zeta |R|}\cdot z_{V,\vec R}\cdot z_{H,{\vec R}}\cdot z_{H^{\dagger}, {\vec R}}.
$$
The $\ell$ hypermultiplet contributions $z_{H^{\dagger}, {\vec R}}$ each contain
$N_{R_a\varnothing}(v e_a/f_a)$, as a factor. Restricting this to \eqref{rhc} we get $N_{R_a\varnothing}(t^{N_a})$, which, as one can show\footnote{See \cite{AHKS} for a proof, and \cite{Awata:2008ed, DHG} for earlier work making use of this.} vanishes if $R_a$ has more than $N_a$ rows. So at this point,
$I^{5d}_{\vec R}$
is non-zero only for those $\ell$-touples of Young diagrams
${\vec R} = (R_1, \ldots, R_a, \ldots R_{\ell})$ for which $R_a$ has no more that $N_a$ rows, for each $a$ between $1$ and $\ell$. Thus, the non-zero fixed point contributions to the instanton sum are the same as the poles of the 3d partition function.
Not only does the sum over Young diagrams truncate, but moreover one can prove that the value of the summand in the instanton partition function is exactly $I^{3d}_{\vec R}$:
$$
I^{3d}_{\vec R}(N, f)= I^{5d}_{\vec R}(e, f),
$$
with identifications
$$e_a/ f_{a} = t^{N_a}/ v.$$
Recall we let $f_{a}  = f_{+, a}$ and $f_{a+\ell}=f_{-, a}$ for $a$ running from $1$ to $\ell$. Finally, we have $q^{\zeta} = q^{\alpha_0}q$.

The vector multiplet contributions in 5d are related to vector multiplet contributions in 3d, and the 5d hypermultiplets to 3d flavors and the instanton counting parameter in 5d to FI term contributions to the potential in 3d. The 5d partition function is actually a product of the instanton sum $I^{5d}_{\vec R}$ together with the perturbative and the one loop factors contained in $r_{5d}$. This equals the partition function of the 5d gauge theory at the root of the Higgs and Coulomb branches in the absence of vortices.  On the 3d gauge theory side, one can prove that this is accounted by the product of $r_{3d}$,
the residue at the $y= y_{\vec \varnothing}$ pole, together with a contribution that is not captured by the theory on the vortex -- this is the partition function of the bulk gauge theory, at the root of the Higgs branch in the absence of vortices.
(From the string theory perspective, this contribution is the partition function of $Y_S$ without branes). One can prove that, taking this into account, the full partition functions on the two sides of the duality are equal.

We have thus proven our main claim \eqref{main} for the case the Gaiotto curve $C$ has genus zero with arbitrary number of punctures. It is elementary to extend this to the case when $C$ is a genus one curve, with arbitrary punctures. We expect the triality to generalize to the case when the Liouville CFT gets replaced by $ADE$ type Toda CFT. The generalization to $A_n$ case will be presented in \cite{AHS}.

\bibliographystyle{utcaps}	
\bibliography{myrefs2}	

\end{document}